\documentclass[aps,prd,reprint,superscriptaddress,nofootinbib]{revtex4-2}

\usepackage{amssymb,amsmath,bm,natbib}
\usepackage[dvipsnames]{xcolor}
\usepackage[colorlinks = true,
            linkcolor = Maroon,
            urlcolor  = Maroon,
            citecolor = Maroon]{hyperref}
\usepackage{microtype}
\usepackage{graphicx}
\usepackage{xspace}
\usepackage[normalem]{ulem}
\usepackage{dsfont}
\usepackage{slashed}

\newcommand{\TT}{\ensuremath{{^3\text{H}}}\xspace}
\newcommand{\He}{\ensuremath{{^3\text{He}^+}}\xspace}

\begin{document}

\title{Relativistic corrections to polarized tritium $\beta$-decay}

\author{Gianluca~Cavoto}
\email{gianluca.cavoto@uniroma1.it}
\affiliation{Dipartimento di Fisica, Sapienza Universit\`a di Roma, Piazzale Aldo Moro 2, I-00185 Rome, Italy}
\affiliation{INFN Sezione di Roma, Piazzale Aldo Moro 2, I-00185 Rome, Italy}

\author{Angelo~Esposito}
\email{angelo.esposito@uniroma1.it}
\affiliation{Dipartimento di Fisica, Sapienza Universit\`a di Roma, Piazzale Aldo Moro 2, I-00185 Rome, Italy}
\affiliation{INFN Sezione di Roma, Piazzale Aldo Moro 2, I-00185 Rome, Italy}
\affiliation{School of Natural Sciences, Institute for Advanced Study, Princeton, NJ 08540, USA}

\author{Guglielmo~Papiri}
\email{gp343@cornell.edu}
\affiliation{Dipartimento di Fisica, Sapienza Universit\`a di Roma, Piazzale Aldo Moro 2, I-00185 Rome, Italy}
\affiliation{Department of Physics, Cornell University, Ithaca, NY 14853, USA}

\author{Antonio~D.~Polosa}
\email{antoniodavide.polosa@uniroma1.it}
\affiliation{Dipartimento di Fisica, Sapienza Universit\`a di Roma, Piazzale Aldo Moro 2, I-00185 Rome, Italy}
\affiliation{INFN Sezione di Roma, Piazzale Aldo Moro 2, I-00185 Rome, Italy}

\date{\today}

\begin{abstract}
\noindent Forthcoming experiments such as Project8 and Ptolemy aim at investigating
with high precision the end-point of the tritium $\beta$-decay spectrum sensitive to the neutrino mass. In light of this, using the standard parametrization in terms of nuclear polar form factors, we analyze the complete relativistic expression for the spectrum of the $\beta$-electron emitted by a tritium nucleus. Given the small parameters in the problem, we systematically discuss the approximations that can be made, and present the first two corrections to the standard lowest order formula.
We particularly discuss the case of an initially polarized target, and the consequences on the spectrum as a function of the neutrino mass. We show that, while it induces an angular anisotropy that can be measured by future experiments, such anisotropy cannot be used as an additional handle to constrain the neutrino mass.
\end{abstract}

\maketitle

%%%%%%%%%%%%%%%%%%%%%%%%%%%%%%%%%%%%%%%%%%%%%%%%%%%%%%%%%%%

\section{Introduction}

The existence of a non-zero neutrino mass is now well established~\cite{Super-Kamiokande:1998kpq,SNO:2001kpb,SNO:2002tuh,deSalas:2020pgw,Esteban:2020cvm,Capozzi:2021fjo}, and provides a compelling evidence for physics beyond the Standard Model.
The differences between the squared masses of the three neutrinos is precisely measured~\cite{deSalas:2017kay,Capozzi:2018ubv,Capozzi:2017ipn}, whereas the value of their absolute mass scale, their mass ordering, as well as their Dirac or Majorana nature are still unknown.

To this end, a number of existing and future experiments aim at determining the lightest neutrino mass. Specifically, this can be done by studying the spectrum of the electron emitted in $\beta$-decay close to its maximum allowed energy, the so-called end-point. Among these, the existing KATRIN~\cite{KATRIN:2001ttj,KATRIN:2019yun,KATRIN:2021uub} and the forthcoming Project~8~\cite{Project8:2014ivu,Project8:2017nal,Project8:2022wqh} and Ptolemy~\cite{Betti:2018bjv,PTOLEMY:2019hkd,Apponi:2021hdu} experiments focus on  the decay of tritium:
\begin{align} \label{eq:decay}
    \TT \to \He + e^- + \bar{\nu}_e\,.
\end{align}
The current best laboratory bound on the effective neutrino mass, $m_\nu^2 \equiv \sum_i |U_{ei}|^2 m_i^2$, is given by $m_\nu \leq 0.8$~eV, as set by KATRIN~\cite{KATRIN:2019yun,KATRIN:2021uub}. Here $m_i$ are the neutrino mass eigenvalues and $U_{ei}$ the entries of the PMNS matrix~\cite[e.g.,][]{Lesgourgues:2013sjj}.
Ptolemy aims at the largest event rate, thanks to a high loading of tritium on, possibly, a flat graphene substrate, as well as an energy resolution as small as $100$~meV~\cite{PTOLEMY:2019hkd,Apponi:2021hdu}. Moreover, the Ptolemy experiment aims at hunting for the cosmic neutrino background via its capture process:
\begin{align}
    \nu_e + \TT \to \He + e^- \,.
\end{align}
In this context, it has been argued that important information about its anisotropy could be obtained from an initially polarized tritium target~\cite{Lisanti:2014pqa}. In order to do that, a much larger target mass than what is currently planned is necessary.

In light of this, it is important to determine the expected spectrum with a theory uncertainty that is better than the experimental precision. In this work we analyze the relativistic corrections affecting the kinematics and nuclear matrix element. For the $\beta$-decay part, such a calculation has already been discussed in some previous works~\cite{Wu:1983sz,Masood:2007rc,Simkovic:2007yi}, but the expression for the spectrum reported completely and without specific assumptions has been presented only in~\cite{Ludl:2016ane}. Moreover, in light of the increasing experimental precision, it is crucial to understand the corrections to the lowest order formula, which is usually employed. 
In this work we discuss the systematic aspects of such corrections in presence of an initially polarized nucleus. We show that, given the expected event rate in Ptolemy, the statistical uncertainty is far smaller than the theoretical error associated to the standard lowest order formula, which thus needs to be improved. 
Specifically, we argue that, for neutrino masses $m_\nu \gtrsim 100$~meV, the leading and next-to-leading order formulae have different shapes, potentially distinguishable by an experiment such as Ptolemy. Moreover, we quantify the effect of initial tritium polarization on the direction of the outgoing electron. Specifically, the spectrum becomes anisotropic, an effect that will be visible within the Ptolemy setup. Nonetheless, such anisotropy is not sufficiently sensitive to the neutrino mass to be used as an additional way to measure it (for a discussion on new physics effects see also~\cite{Canning:2022nye}).

%%%%%%%%%%%%%%%%%%%%%%%%%%%%%%%%%%%%%%%%%%%%%%%%%%%%%%%%%%%

\section{Review of the calculation}

In this work we follow the approach outlined in~\cite{Simkovic:2007yi}, where the relativistic matrix element is obtained from an hadronic model for the tritium nucleus~\cite{Kim:1965zza}:
\begin{align} \label{eq:formfactors}
    \begin{split}
        \mathcal{M}={}&\frac{G_F V_{ud}}{\sqrt{2}}\, \bar{u}(P_e)\gamma_{\mu}(1-\gamma_5 )v(P_{\nu}) \\
        &\times \bar{u}(P_f)\bigg[G_V \big(q^2\big)\gamma^{\mu} + i\frac{G_M \big(q^2\big)}{2 M_i}\sigma^{\mu \nu}q_{\nu} \\
        &\quad- G_A\big(q^2\big)\gamma^{\mu}\gamma_5 - G_P\big(q^2\big)q^{\mu}\gamma_5\bigg]u(P_i)\,,  \end{split}
\end{align}
where $G_F$ is the Fermi constant and $V_{ud}$ the entry of the CKM matrix.
Moreover, $P_i$ and $P_f$ are the four-momenta of \TT and \He (with masses $M_i$ and $M_f$), $P_e$ and $P_\nu$ are those of the electron and antineutrino, and $q\equiv P_f-P_i$ is the momentum transferred to the nucleus. The form factors are~\cite{Simkovic:2007yi}
\begin{align}
    \begin{split}
        & G_V = \frac{g_V}{\left(1-q^2/M_V^2\right)^2}\,, \quad G_M = \frac{g_M}{\left(1-q^2/M_V^2\right)^2}\,, \\
        & G_A = \frac{g_A}{\left(1-q^2/M_A^2\right)^2}\,, \quad G_P = \frac{2M_i}{m_\pi^2-q^2}G_A\,,
    \end{split}
\end{align}
with $m_\pi$ the pion mass and $M_{A,V}$ two mass scales of the order of the GeV. The value of the couplings $g$ in the equations above can be found in Table~\ref{tab:numbers}.

It should be noted that the matrix element in Eq.~\eqref{eq:formfactors} does not include the effects of three-body interactions within the tritium nucleus, which can instead play some role~\cite[e.g.,][]{Kamada:2001tv,Pieper:2007ax,Doi:2011gq}. The inclusion of these effect is, however, rather non-trivial, and it goes beyond the scope of the present work.

A version of the full relativistic spectrum has also been discussed in~\cite{Masood:2007rc}. However, the matrix element employed there is not computed from first principles, but rather parametrized using Lorentz invariance together with some considerations regarding which terms should be the leading ones. While this approach reproduces well the lowest order formula, it is not the most suitable one for computing higher order corrections.

In the following we neglect the momentum dependence of the form factors. This approximation is correct up to relative corrections given by
\begin{align}
    \begin{split}
        \frac{q^2}{M_{A,V}^2} \simeq \frac{m_e^2}{M_{A,V}^2} \sim 10^{-7} \quad &\text{for } G_V, G_M \text{ and } G_A\,, \\
        \frac{q^2}{m_\pi^2} \simeq \frac{m_e^2}{m_\pi^2} \sim 10^{-5} \quad &\text{for } G_P\,.
    \end{split}
\end{align}

Given the above, one can compute the fully relativistic decay rate. In particular, we consider a tritium nucleus initially polarized along a direction $\hat{\bm{n}}$. Possible ways to achieve a high polarization fraction have been discussed in~\cite{Lisanti:2014pqa}. In this case, the four-spinors in the laboratory frame obey,
\begin{align}
    u(P_i)\bar u(P_i) = M_i\left(\mathds{1} + \gamma^0 \right) \frac{\mathds{1} + \gamma^5 \bm\gamma\cdot{\hat{\bm n}}}{2} \,.
\end{align}
The corrections due to tritium polarization are expected to be dependent 
on electron velocity~\cite{weisskopf} and, indeed, will vanish in the non-relativistic limit.
Summing over the polarizations of the final states, the rate in the laboratory frame is given by,
\begin{align} \label{eq:dGdEimplicit}
    \frac{d\Gamma}{dE_e d\Omega} ={}& \frac{{G_F^2 V_{ud}^2}}{8\pi^4} F(Z,E_e)\frac{M_i^2 p_e}{M_{12}^2} \sqrt{\Delta\left(\Delta-2m_{\nu}\frac{M_f}{M_i}\right)} \notag \\
    \begin{split}
        & \times \Big[g_V^2 \mathcal{W}_{VV} + g_A^2 \mathcal{W}_{AA} + g_A g_V \mathcal{W}_{AV} \\
        & \quad\;\; + g_A^2 \mathcal{W}_{AP} + g_A^2 \mathcal{W}_{P\!P} + g_V g_M \mathcal{W}_{V\!M} 
    \end{split} \\
    & \quad\;\; + g_A g_M \mathcal{W}_{AM}  + g_M^2 \mathcal{W}_{M\!M} \Big] \notag\,,
\end{align}
where $E_e$ is the electron energy and $\Omega$ its emission angle, defined with respect to the direction of the tritium polarization. Moreover, $M_{12}^2\equiv M_i^2+m_e^2-2M_i E_e$, and $F(Z,E_e)$ is the Fermi factor accounting for the electromagnetic interaction between the outgoing electron and the decayed nucleus. This is given by~\cite[e.g.,][]{Blatt:1952ije},
\begin{align}
    F(Z,E_e) = 2(1+\gamma)\left( 2p_e R\right)^{-2(1-\gamma)} e^{\pi y} \frac{\left|\Gamma(1+i y)\right|^2}{\Gamma(2\gamma+1)^2} \,,
\end{align}
with $\gamma \equiv \sqrt{1 - \alpha^2 Z^2}$, $y = \alpha Z E_e/p_e$. Moreover, $Z=2$ and $R=2.8840 
\times 10^{-3}/m_e$ are, respectively, the atomic number and radius of \He~\cite{elton1958semi,Mertens:2014nha,Ludl:2016ane}. Finally, we define $\Delta\equiv E_e - E_e^\text{max}$, with the relativistic end-point given by\footnote{We note a typo in the denominator of Eq.~(7) in~\cite{Simkovic:2007yi}, where $M_f$ should be replaced with $M_i$.}
\begin{align} \label{eq:Emax}
    E_e^\text{max} = \frac{M_i^2 + m_e^2 - \left( M_f + m_\nu \right)^2}{2M_i}\,.
\end{align}
The $\mathcal{W}$'s appearing in Eq.~\eqref{eq:dGdEimplicit} result from summing the squared matrix element over final polarizations and integrating it over the relativistic three-body phase space, as outlined in~\cite{Simkovic:2007yi}. Their expressions (again, in the approximation of constant form factors) are rather cumbersome, and we will not report them here.\footnote{The complete expression for the matrix elements and the $\mathcal{W}$ factors can be provided in \textsc{FORM} and \textsc{Wolfram Mathematica}\textsuperscript{\textregistered} format upon request.}

%%%%%%%%%%%%%%%%%%%%%%%%%%%%%%%%%%%%%%%%%%%%%%%%%%%%%%%%%%%

\section{Expansion in the small parameters} \label{sec:systematic}

The dimensionful quantities entering the $\mathcal{W}$ terms come with different sizes, resulting in a number of small parameters that one can use to  approximate the full expression, Eq.~\eqref{eq:dGdEimplicit}. These parameters are chosen to be (see also~\cite{Ludl:2016ane}),
\begin{align} \label{eq:smallparam}
    \begin{split}
        \sqrt{\frac{Q}{m_e}} \sim 0.1 \,, \quad \frac{m_e}{m_\pi} \sim{}& 10^{-3} \,, \quad  \frac{m_e}{M_i} \sim 10^{-4}\,, \\
        \frac{Q}{M_i} \sim \frac{\sqrt{Q m_e}}{M_i} \sim 10^{-5}  \,,& \quad
        \frac{|\Delta|}{M_i} \sim \frac{m_\nu}{M_i} \sim 10^{-10}\,, 
    \end{split}
\end{align}
where $Q\equiv M_i - M_f - m_e$ is the $Q$-value. Since the spectrum is most sensitive to the neutrino mass close to the end-point, we also set $|\Delta|\sim m_\nu$, and assumed conservatively $m_\nu \sim 1$~eV. For everything else, we used the numerical values in Table~\ref{tab:numbers}.
\begin{table*}[t]
    \centering
    \begin{tabular}{c|c|c|c|c|c|c}
        $M_i$ [eV]~\cite{Myers:2015lca} & $m_\pi$ [eV]~\cite{ParticleDataGroup:2020ssz} & $m_e$ [eV]~\cite{ParticleDataGroup:2020ssz} & $Q$ [eV]~\cite{Myers:2015lca} & \;$g_V$\; & $g_A$~\cite{Markisch:2018ndu} & $g_M$~\cite{Simkovic:2007yi} \\ 
        \hline\hline
         $2.80943249663(19) \times 10^9$ & $137.27360(27) \times 10^{6}$ & $0.5109989461(31) \times 10^{6}$ & $18.59201(7) \times 10^{3}$ & $1$ & $1.27641(55)$ & $-6.106$
    \end{tabular}
    \caption{Values used in this work. In order to keep track of all the uncertainties, we also report the experimental ones. For the pion mass we report the average between charged and neutral states. The value $g_V=1$ is implied by the conserved vector current hypothesis, while, to the best of our knowledge, the uncertainty on $g_M$ is not found in the literature. We also note that the value of $g_A$ can also be extracted directly from the tritium half-life, obtaining $g_A = 1.247$~\cite{Budick:1991zb,Simkovic:2007yi,Simpson:1987zz}. We checked that this different choice does not lead to appreciable differences in our analysis.}
    \label{tab:numbers}
\end{table*}
We also report the current errors on the measured parameters. In fact, while this does not affect the present work, for higher order corrections, the uncertainties coming from the parameters can compete with the theoretical uncertainties. This is especially true for the masses, the $Q$-value and the $g$ couplings. The errors on overall factors are, instead, expected not to play a role, since a standard experimental analysis would fit the overall normalization anyway. In light of this, the precise values of $G_F$ and $V_{ud}$ and of their uncertainties are irrelevant for the present analysis.

Following the expansion scheme outlined above, one can systematically expand the $
\mathcal{W}$'s. In particular, we retain all corrections up to $\mathcal{O}\big(10^{-6}\big)$, which are relevant for Ptolemy at maximum coverage (see Section~\ref{sec:ptolemy}). In particular, one finds, 
\begin{widetext}
    \begin{subequations} \label{eq:scalings}
        \begin{align} 
            \mathcal{W}_{VV} ={}& E_e \left[ (-\Delta + m_{\nu}) - m_{\nu} \frac{m_e}{M_i} - m_{\nu}\frac{Q}{M_i} \right] + \mathcal{O}\left(\frac{m_e^2 Q \Delta}{M_i^2}\right) \,, \\
            \mathcal{W}_{AA} ={}& E_e \left[3 (-\Delta+m_{\nu})-3 m_{\nu}\frac{m_e}{M_i} + (-4\Delta+m_{\nu})\frac{Q}{M_i} \right] - \sqrt{2} E_e \cos\theta \bigg[ 2(\Delta-m_{\nu})\sqrt{\frac{Q}{m_e}} + (-\Delta-m_{\nu})\frac{\sqrt{Q m_e}}{M_i} \notag \\
            & - \frac{3}{{2}}(-\Delta+m_{\nu})\left(\frac{Q}{m_e}\right)^{3/2} + \frac{23}{16} (-\Delta+m_{\nu})\left(\frac{Q}{m_e}\right)^{5/2} \bigg] + \mathcal{O}\left(m_e \Delta \left(\frac{Q}{m_e}\right)^{7/2}\right) \,, \\
            \begin{split}
                \mathcal{W}_{AV} ={}& 4E_e(-\Delta+m_{\nu})\frac{Q}{M_i} - \sqrt{2} E_e \cos\theta \bigg[ - 2(-\Delta+m_{\nu})\sqrt{\frac{Q}{m_e}} + (-3\Delta+5m_{\nu})\frac{\sqrt{Q m_e}}{M_i} \\
                & + \frac{3}{{2}} (-\Delta+m_{\nu}) \left(\frac{Q}{m_e}\right)^{3/2} - \frac{23}{16} (-\Delta+m_{\nu})\left(\frac{Q}{m_e}\right)^{5/2} \bigg] + \mathcal{O}\left(m_e \Delta \left(\frac{Q}{m_e}\right)^{7/2}\right) \,,
            \end{split} \\
            \mathcal{W}_{AM} ={}& -4E_e(-\Delta+m_{\nu})\frac{Q}{M_i} +2 \sqrt{2}\cos\theta E_e (-\Delta+m_{\nu})\frac{\sqrt{Q m_e}}{M_i} + \mathcal{O}\left(m_e \Delta \frac{Q}{M_i}\sqrt{\frac{Q}{m_e}}\right) \,, \\
            \mathcal{W}_{VP} ={}& +2 \sqrt{2}\cos\theta E_e (-\Delta+m_{\nu})\frac{m_e}{m_{\pi}}\frac{\sqrt{Q m_e}}{m_{\pi}} + \mathcal{O}\left(m_e \Delta \left(\frac{Q}{m_{\pi}}\right)^{3/2} \sqrt{\frac{m_e}{m_\pi}}\right) \,,
        \end{align}
    \end{subequations}
\end{widetext}
with all the others being subleading,
\begin{subequations}
    \begin{align}
        \mathcal{W}_{VM} ={}& \mathcal{O}\left( m_e \Delta  \left(\frac{Q}{M_i}\right)^{3/2}\sqrt{\frac{m_e}{M_i}} \right) \,, \\
        \mathcal{W}_{MM} ={}& \mathcal{O}\left( m_e \Delta  \left(\frac{Q}{M_i}\right)^{3/2}\sqrt{\frac{m_e}{M_i}} \right) \,, \\
        \mathcal{W}_{MP} ={}&  \mathcal{O}\left( m_e \Delta \left(\frac{m_e}{M_i}\right)^2 \left(\frac{Q}{m_{\pi}}\right)^{3/2} \sqrt{\frac{m_e}{m_\pi}} \right) \,, \\
        \mathcal{W}_{AP} ={}& \mathcal{O}\left( m_e \Delta \left(\frac{Q}{m_{\pi}}\right)^2 \frac{m_e}{M_i} \right) \,, \\
        \mathcal{W}_{PP} ={}& \mathcal{O}\left( m_e \Delta \left(\frac{Q}{m_{\pi}}\right)^2 \left(\frac{m_e}{m_{\pi}}\right)^2 \right) \,.
\end{align}
\end{subequations}
As anticipated, the angular dependence due to the initial polarization vanishes in the non-relativistic limit, $Q \sim p_e \to 0$.

Before computing the decay rate and its leading corrections, we briefly comment about the expansion presented in~\cite{Simkovic:2007yi}. There, in order to estimate the relative sizes of the various terms, it is set $M_f=M_i$. However, this equality holds true only at lowest order, and receives corrections already at order $m_e/M_i$. Because of this, the scalings reported in Eq.~(28) of~\cite{Simkovic:2007yi} lead 
to an \emph{overestimate} of the corrections (in all cases but one). In fact, when one keeps track of the fact that $M_i \neq M_f$, additional cancellations happen, reducing the size of the correction. 
Moreover, we report the presence of a wrong sign in the expression of $\mathcal{W}_{AA,VV,AV}$, given in Eq. (29) of~\cite{Simkovic:2007yi}, for the first term in the parentheses proportional to $(g_V +g_A)^2$.

\begin{figure*}
    \centering
    \includegraphics[height=5.9cm]{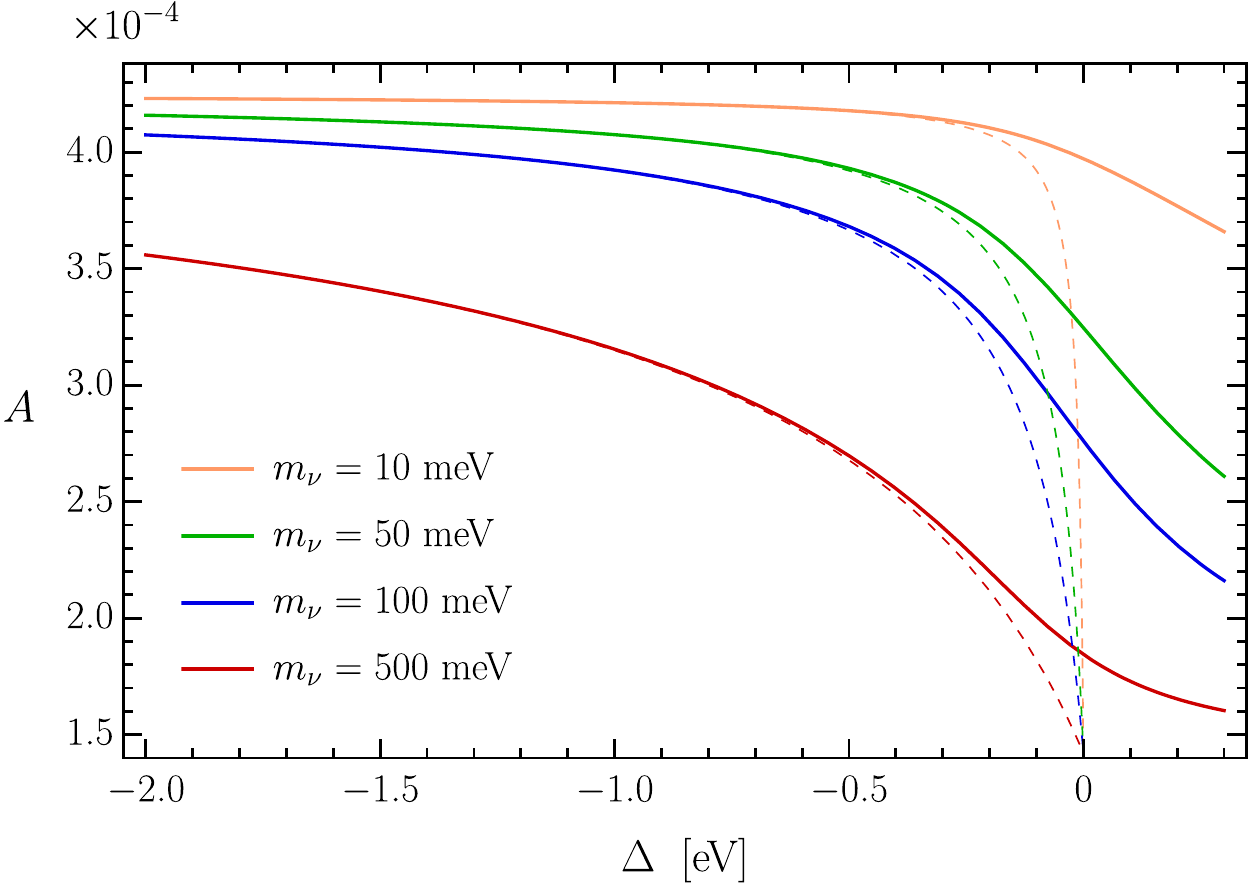} \hspace{0.5em}
    \includegraphics[height=5.95cm]{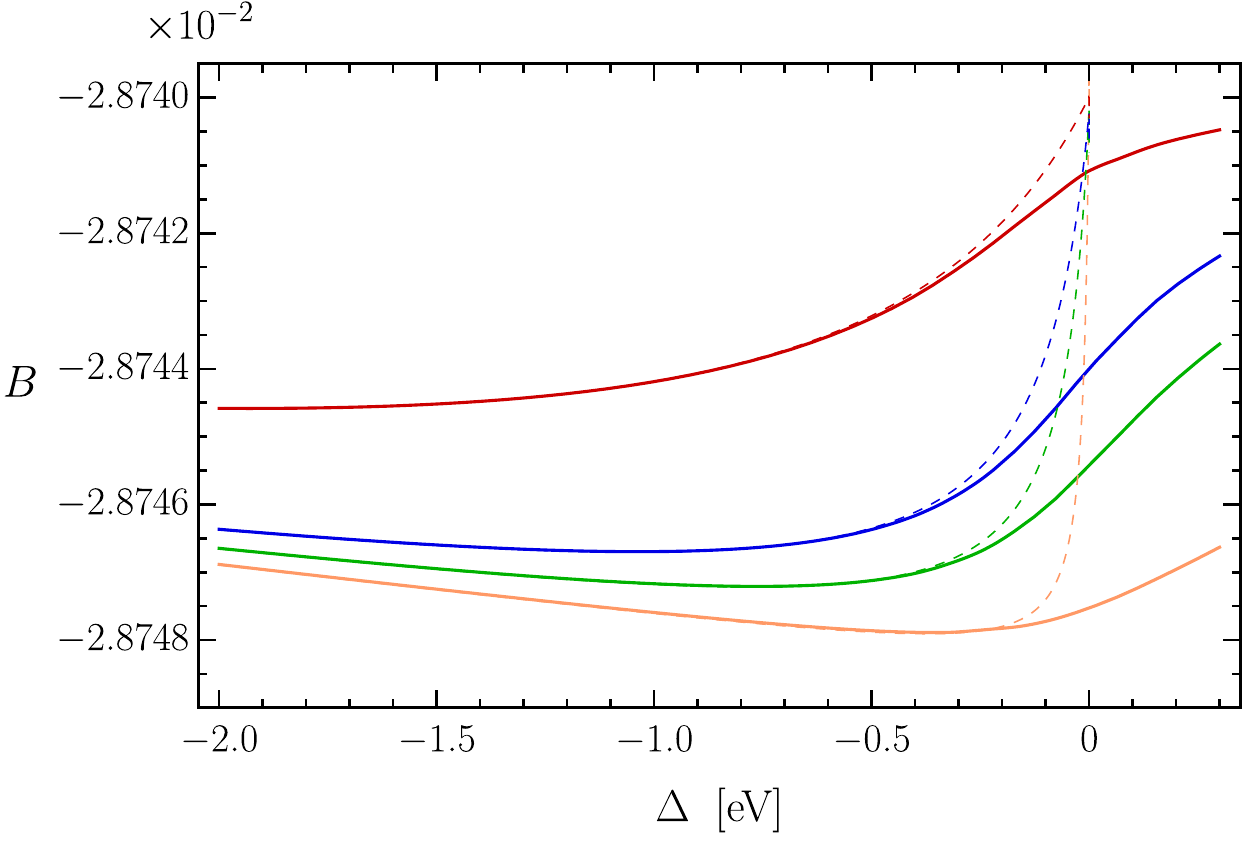}
    \caption{Relative corrections to the leading order rate, as parametrized in Eq.~\eqref{eq:Gammatilde}. Dashed lines show the result in absence of experimental resolution. For a sufficiently heavy neutrino, the difference in shape of the two expression is substantially larger that the expected statistical uncertainty for Ptolemy at maximum loading. For each curve, the end-point, $\Delta = 0$, is computed using the exact relativistic expression evaluated at the relevant value of the neutrino mass. The non-zero values for $\Delta \geq 0$ are due to the experimental resolution in Eq.~\eqref{eq:res}, which pushes some events above the end-point.}
    \label{fig:rates}
\end{figure*}

We observe that, from Eqs.~\eqref{eq:scalings}, the leading order expression for the differential rate reads
\begin{align} \label{eq:LO}
    \begin{split}
        \frac{d\Gamma_{(0)}}{dE_e} ={}& \frac{\left( G_F V_{ud} \right)^2}{2\pi^3} F(Z,E_e)\left(g_V^2+3g_A^2\right) \\
        & \times E_e p_e (m_\nu -\Delta) \sqrt{\Delta(\Delta - 2m_\nu)} \,,    
    \end{split}
\end{align}
in agreement with the literature~\citep[e.g,][]{Masood:2007rc,Simkovic:2007yi,PTOLEMY:2019hkd}.
The leading and next-to-leading order expressions for the electron end-point are, instead,
\begin{subequations}
    \begin{align}
        E_{e,(0)}^{\rm max} ={}& m_e + Q - m_\nu \,, \\
        E_{e,(1)}^{\rm max} ={}& m_e + Q - m_\nu - m_e \frac{Q}{M_i} \,.
    \end{align}    
\end{subequations}

%%%%%%%%%%%%%%%%%%%%%%%%%%%%%%%%%%%%%%%%%%%%%%%%%%%%%%%%%%%

\section{Relevance for Ptolemy} \label{sec:ptolemy}

Thanks to the storage of atomic tritium on a solid state substrate, Ptolemy aims at employing a large mass of tritium, possibly up to $100$~g~\cite{PTOLEMY:2019hkd}, which would also allow to hunt for the cosmic neutrino background. 

\begin{figure}
    \centering
    \includegraphics[width=\columnwidth]{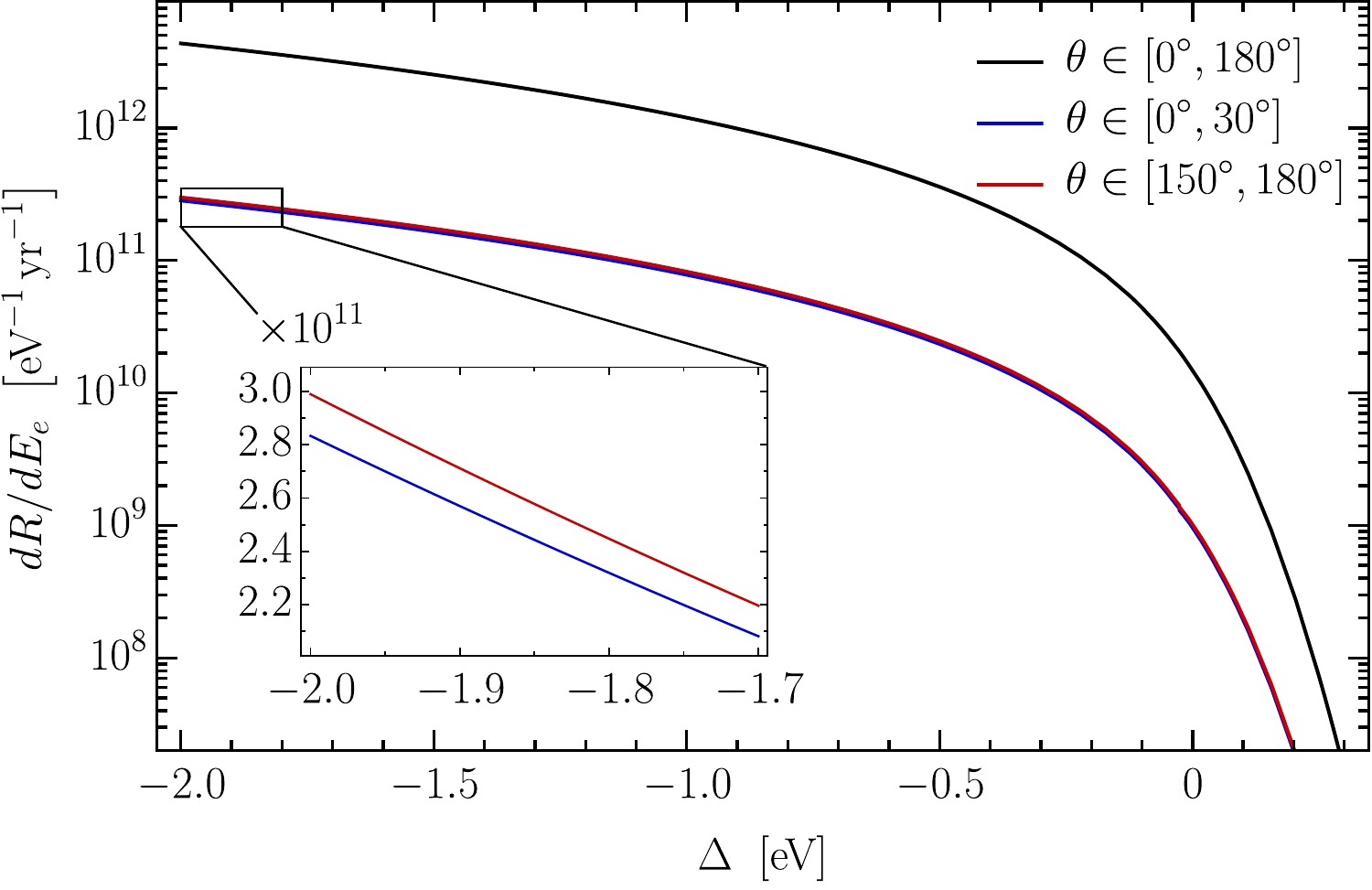}
    \caption{Event rate as a function of the electron energy, $E_e$, obtained integrating Eq.~\eqref{eq:Gammatilde} over different ranges for the polar angle, $\theta$. The neutrino mass is set to $m_\nu = 100$~meV. The inset highlights the asymmetry related to the emission of the electron in opposite directions, $\theta\in[0^{\circ},30^{\circ}]$ and $\theta\in[150^{\circ},180^{\circ}]$. The event rate is computed as $R = N_{\rm T} \tilde\Gamma$, where $N_{\rm T} \simeq 2 \times 10^{25}$ is the number of nuclei corresponding to a mass $m_{\rm T} = 100$~g.}
    \label{fig:diff_rate_pol}
\end{figure}

From Eqs.~\eqref{eq:dGdEimplicit} and \eqref{eq:scalings} we see that the rate can be parametrized as
\begin{align} \label{eq:Gammatilde}
    \frac{d\tilde\Gamma}{dE_e d\Omega} \equiv \frac{1}{4\pi} \frac{d \tilde \Gamma_{(0)}}{dE_e} \big( 1 + A(E_e) + B(E_e) \cos\theta\big) \,,
\end{align}
where $\tilde\Gamma$ is the theoretical rate convoluted with the experimental resolution, taken to be a Gaussian of width $\sigma = 100$~meV:
\begin{align} \label{eq:res}
    \frac{d\tilde\Gamma}{dE_e d\Omega} = \int_{-\infty}^{\infty} dE^\prime \, \frac{\text{e}^{-(E_e - E^\prime)/(2 \sigma^2)}}{\sqrt{2\pi} \sigma} \, \frac{d \Gamma}{dE^\prime d\Omega} \,.
\end{align}
The unpolarized rate is simply obtained by setting $\cos\theta \to 0$ in Eq.~\eqref{eq:Gammatilde}.
Importantly, since in a realistic analysis one would also fit the position of the end-point~\cite{PTOLEMY:2019hkd}, in the following results we use the expression for the exact relativistic $E_e^{\rm max}$, Eq.~\eqref{eq:Emax}, evaluated at the value of the neutrino mass under consideration.
In Figure~\ref{fig:rates} we report the behavior of $A$ and $B$. In Figure~\ref{fig:diff_rate_pol} we, instead, report the number of events expected for $m_{\rm T}=100$~g of tritium, as a function of the electron energy  and for different directions of emission, defined with respect to the initial polarization.

Considering (conservatively) events in the range $\Delta \in [-2,0]$~eV, Eq.~\eqref{eq:LO} returns an expectation of $N\sim 10^{12}$~events/year. Assuming a Poisson distribution, this corresponds to a relative statistical uncertainty of order $10^{-6}$ after a year of data taking. As one can see from Figure~\ref{fig:rates}, for sufficiently heavy neutrinos, this is considerably smaller than the relative difference introduced by the next-to-leading order correction to the rate. The shape of the latter, in fact, varies by a fraction of the order of $10^{-4} - 10^{-5}$ over a large range of energies. Such corrections must then be included in order to avoid theoretical biases on the parameter estimation.
Similarly, the relative directional asymmetry is of the order of $5\%$, with more events expected in the direction anti-parallel to the initial polarization. This effect should be visible within the Ptolemy setup.

\begin{figure}
    \centering
    \includegraphics[width=\columnwidth]{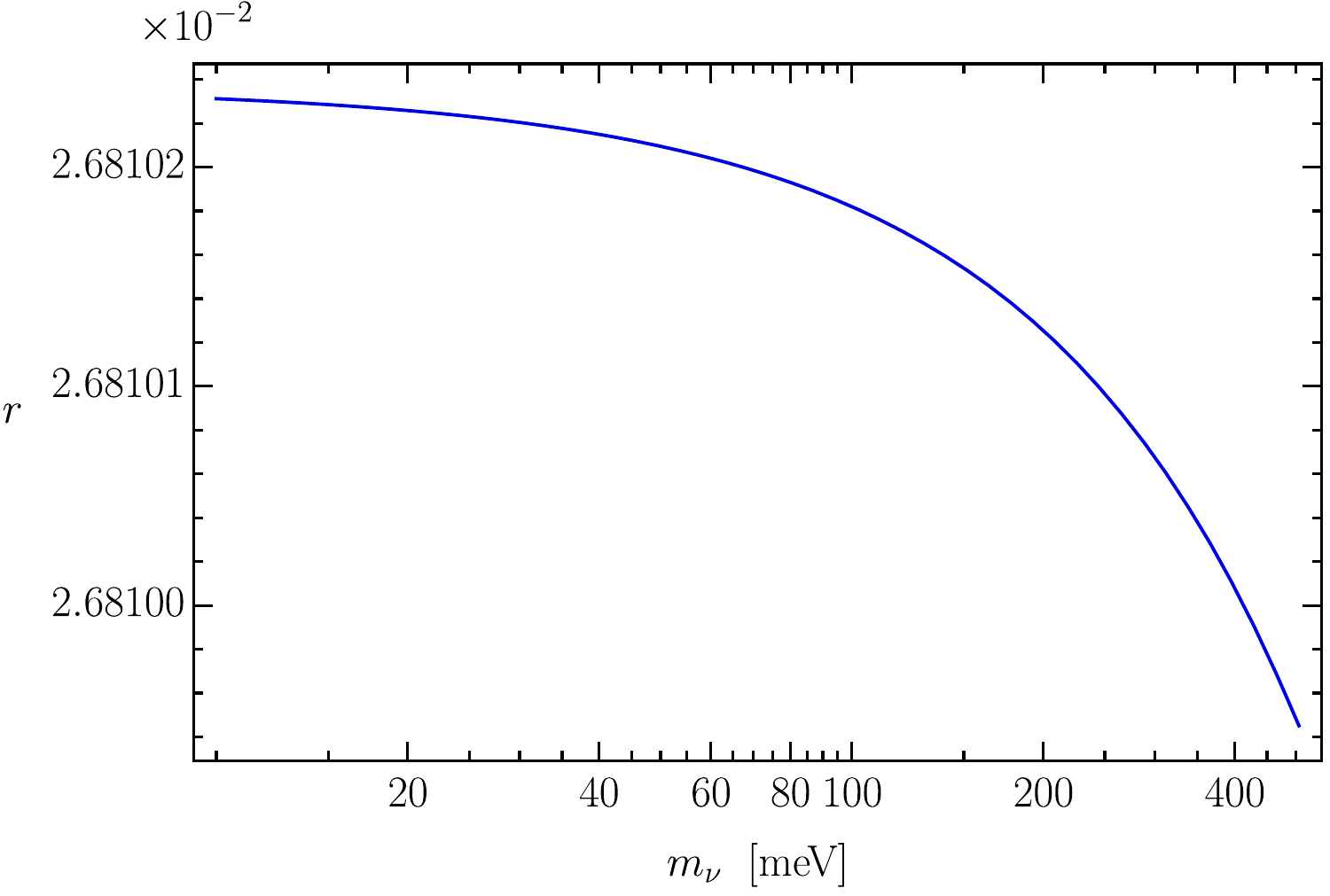}
    \caption{Asymmetry ratio, as defined in Eq.~\eqref{eq:r}, as a function of the neutrino mass. The relative variation is comparable to the expected statistical error for Ptolemy at maximum coverage.}
    \label{fig:r}
\end{figure}

The degree of asymmetry depends on the value of the neutrino mass. To quantify this, we define the ratio,
\begin{align} \label{eq:r}
    r \equiv \frac{N_- - N_+}{N_- + N_+} \,,
\end{align}
where $N_+$ is the number of electrons emitted parallel to the initial tritium polarization, with an angle $\theta \in [0^\circ,30^\circ]$, while $N_-$ are those emitted anti-parallel, with an angle $\theta \in [150^\circ, 180^\circ]$. The events are counted in an energy bin such that $\Delta \geq -2$~eV, and for $m_{\rm T} = 100$~g. As one can see from Figure~\ref{fig:r}, with varying neutrino mass, the asymmetry ratio exhibits a relative variation of order $10^{-5}$, which is an order of magnitude smaller than what is na\"ively expected from the behavior of $B$ (Figure~\ref{fig:rates}). Moreover, the relative error expected for a small anysotropy, $r\ll 1$, is given by,
\begin{align}
    \frac{\sigma_r}{r} \simeq \frac{1}{r}\sqrt{\frac{1}{2N}} \,.
\end{align}
For Ptolemy at maximum coverage, $N \sim 10^{12}$, one then expects $\sigma_r/r \sim 10^{-5}$. The observation of the dependence in Figure~\ref{fig:r} is thus difficult.

%%%%%%%%%%%%%%%%%%%%%%%%%%%%%%%%%%%%%%%%%%%%%%%%%%%%%%%%%%%

\section{Conclusion}

In the coming years a host of existing and future experiments should place increasingly more stringent bounds on the neutrino mass. Among these, Ptolemy might reach unprecedented rates, and correspondingly low statistical uncertainties. 
This improvement on the experimental side  demands precise theoretical prediction. 

In this paper we discussed the corrections due to relativistic effects in the nuclear decay of tritium. Specifically, by systematically keeping track of the approximations made, we revisit some results available in the literature, and present a revised compendium of what we identify as the most relevant corrections to the $\beta$-decay spectrum for the experimental setup as envisioned by Ptolemy. We show that, given the expected number of events, these corrections are important to avoid theory biases.

Moreover, due to the presence of an external magnetic field in the Ptolemy electromagnetic filter, the initial tritium is expected to be at least partially polarized. We discuss the consequence of this on the electron spectrum, and show how the expected anisotropy likely will not provide a further handle to extract the neutrino mass. Together with the consideration made in~\cite{Lisanti:2014pqa}, this makes the initial polarization of the tritium nuclei an information that can hardly be employed by forthcoming experiments.

This is  a first step towards the development of theory predictions, which are accurate enough to face the forthcoming experimental challenges. 
In particular, we note that experiments such as the ones mentioned above use (or plan to use) tritium bound either in a molecule or to a solid state substrate. The presence of a spectator system comes with a number of difficulties and further corrections, which have been discussed in~\cite{Saenz:1997zz,Saenz:1997zza,Saenz:2000dul,Bodine:2015sma,Cheipesh:2021fmg,Nussinov:2021zrj,Tan:2022eke,PTOLEMY:2022ldz} (see also~\cite{Mikulenko:2021ydo,Brdar:2022wuv}).
Consequently, several more effects will then need to be taken into account to achieve the required precision. In particular, atomic, molecular and solid state ones as, for example, possible excitation of and screening from orbital electrons, as well as spatial localization effects and the excitation of nuclear degrees of freedom. In line with the present study, it would be interesting to understand the implications of an external magnetic field on the binding between tritium and different solid state substrates.

%%%%%%%%%%%%%%%%%%%%%%%%%%%%%%%%%%%%%%%%%%%%%%%%%%%%%%%%%%%

\begin{acknowledgments}
We thank L.~E.~Marcucci and M.~Viviani for discussions on the parametrization of the nuclear matrix element, and D.~del Re, S.~Gariazzo and A.~Messina for discussions about statistical analyses and the bounds on the neutrino mass. We are also grateful to W.~Rodejohann for important feedback on the manuscript.
For most of the development of this work A.E. has been a Roger Dashen Member at the Institute for Advanced Study, whose work was also supported by the U.S. Department of Energy, Office of Science, Office of High Energy Physics under Award No. DE-SC0009988.
\end{acknowledgments}

\bibliographystyle{apsrev4-1}
\bibliography{biblio}

%%%%%%%%%%%%%%%%%%%%%%%%%%%%%%%%%%%%%%%%%%%%%%%%%%%%%%%%%%%

\end{document}